\documentstyle[eqsecnum,aps]{revtex}

\begin{document}
\title{Isospin structure of one- and two-phonon GDR excitations }
\author{A.F.R. de Toledo Piza$^{(a)}$, M.S. Hussein$^{(a)}$, B. V. Carlson$^{(b)}$,
C.A. Bertulani$^{(c)}$, L.F. Canto$^{(c)}$ and S. Cruz-Barrios$^{(d)}$}
\address{$^{(a)}$Instituto de F\'{\i}sica,Universidade de 
S\~{a}o Paulo,
01498 S\~{a}o Paulo, SP, Brazil\\
$^{(b)}$Departamento de F\'{\i}sica, Instituto Tecnol\'ogico da
Aeron\'autica - CTA, 12228-900 S\~ao Jos\'e dos Campos, SP, Brazil\\
$^{(c)}$Instituto de F\'{\i}sica, Universidade Federal 
do Rio de Janeiro,
21945-970 Rio de Janeiro, RJ, Brazil \\
$^{(d)}$
Departamento de F\'{\i}sica Aplicada, Universidad 
de Sevilla, 41080
Sevilla, Spain }
\date{\today}
\maketitle

\begin{abstract}
Isospin is included in the description of Coulomb excitation of multiple
giant isovector dipole resonances. In the excitation of even-even nuclei,
a relevant portion of the excitation strength is shown to be associated 
with 1$^{+}$ two-phonon states, which tends to be hindered or completely 
supressed in calculations in which the isospin degree of freedom is not
considered.  We find that the excitation cross sections is strongly
dependent on the ground state isospin.
\end{abstract}

\section{Introduction}

Coulomb excitation of two-phonon giant resonances in heavy ion collisions at
relativistic energies was predicted by Baur and Bertulani in 1986 \cite{BB86}
and generated considerable interest during the last few years \cite
{Em94,CF95,Ber98a}. The double giant isovector dipole resonance (DGDR) has
now been observed in several nuclei: $^{136}$Xe \cite{Sc93}, $^{197}$ Au 
\cite{Au93}, and $^{208}$Pb \cite{Ri93,Bee93}. The double giant isoscalar
quadrupole resonance has been also observed in the $^{40}{\rm Ca}\left(
^{40} {\rm Ca},\;^{40}{\rm Ca}\; p\right)$ reaction, at $44\cdot A$ MeV
laboratory energy \cite{Cho84}. Data on the DGDR were confronted with
results of coupled channels Coulomb excitation theory \cite{Ca94,BCH96} and
the major conclusion reached was that theoretical predictions underestimate
the data by a factor 2-3 in the cases of $^{136}$Xe and $^{197}$Au. A
similar discrepancy, albeit appreciably smaller, was found in $^{208}$Pb 
\cite{Bo96}.

Several effects, not taken into account in the coupled channel theory, were
considered to explain this discrepancy. Among these are anharmonicities \cite
{Vo95,BD97} and the Brink-Axel mechanism \cite{BA}, to cite a few. In this
paper we examine the relevance of isospin effects in the excitation process
relying on the pioneering work of Fallieros, in which the isospin splitting
of the isovector giant dipole resonance was first analyzed \cite
{Fa65,Fa70,Fa71}. We next present an extension of this work to the double
giant isovector dipole states \cite{guara}, before turning to more technical
details of the coupled channels calculation.

The relevant dipole excitation operator is the component $T_3=0$ of an
isovector ($T=1$) operator, $D_{T=1,T_3=0}$. Thus, when acting on a target
nucleus with isospin quantum numbers $T$, $T_3=T$, two GDR states will be
generated having isospin quantum numbers $T_I=T$ and $T_I=T+1$ ($T_I=T-1$ is
forbidden due to charge conservation), where the label $I$ has been
introduced to designate the (intermediate) one-phonon state. If the dipole
operator is applied again to these one-phonon states, final states with
isospin $T_f=T,\;T+1$ and $T+2$ will be generated. In order to take into
account the bosonic nature of isovector phonons in these final states, one
must keep track of another isospin quantum number, namely the {\it total}
isospin $\Im$ of the {\it two} dipole phonon operator, which can take the
values $\Im =0,\;1,\;2.$ These values will constrain the total angular
momentum of the two coupled phonons through symmetry requirements. For a
nucleus with $J^\pi = 0^{+}$ ground state, the DGDR may have $J^\pi =
0^{+},1^{+}$ and $2^{+}$, and for the state $1^{+}$ one must have $\Im =1$.
If isospin is not taken into account, $1^{+}$ states, reached from the $%
1^{-} $ GDR, will be quenched \cite{BPV96}, since by itself it is
antisymmetric under exchange of the two phonons. However, $1^{+}$ states
will in general contribute to the excitation cross section, if explicit
reference to its $\Im =1$ nature is made. In this case, the exchange
symmetry is odd both in the spin and isospin spaces, so that the product has
even symmetry, as required.

The energy splitting of the isodoublet GDR was studied in \cite{Fa71} and
was found to be related to the symmetry energy and to the average
particle-hole interaction, leading to the estimate

\begin{equation}
\Delta^{(1)}_{T+1} = E_{T+1}^{(1)}-E_T^{(1)}\cong 60\;\frac{T+1}A\; [{\rm MeV%
}]\;.  \label{fall1}
\end{equation}

\noindent The energy splitting of the isotriplet DGDR can be estimated in a
similar way. Since the $T_f=T+2$ state is a double isospin analog state it
involves twice the displacement energy, and we may write

\begin{equation}
\Delta^{(2)}_{T+2}= E_{T+2}^{(2)}-E_T^{(2)}\cong 120\;\frac{T+2}A\;[{\rm MeV}%
]\;.  \label{fall2}
\end{equation}

\noindent Furthermore, for the $T_f=T+1$ state we may write

\begin{equation}
\Delta^{(2)}_{T+1}= E_{T+1}^{(2)}-E_T^{(2)}\cong 60\;\frac{T+1}A\;[{\rm MeV}%
]\;.  \label{fall3}
\end{equation}

\noindent We give in table I the energies of the isospin doublet and triplet
states in $^{208}Pb$ ($T=22$) as obtained from these expressions. \bigskip

%
%

\begin{center}
\begin{tabular}{lll}
\hline
Energy shift (MeV)\ \ \ \ \ \ \ \ \  & $^{208}$Pb\ ($T=22$)\ \ \ \ \ \ \ \ \
\  & $^{48}$Ca\ ($T=4$) \\ \hline
$\Delta _{T+1}^{(1)}\;$ & 6.63 \ \  & 6.25 \\ 
$\Delta _{T+2}^{(2)}$ & 13.85 & 15 \thinspace \thinspace \\ 
$\Delta _{T+1}^{(2)}$ & 6.63 \ \  & 6.25 \\ \hline
\end{tabular}
\end{center}

\medskip

\begin{center}
{\bf Table I.} {\it Isospin splitting (in MeV) of one- and two-phonon states
in $^{208}Pb$} and {\it $^{48}Ca.$}
\end{center}

\bigskip

\noindent Having given the above account on the isospin structure of the one
and two-phonon dipole states, we now turn to the required modifications of
the coupling interaction for the coupled channels calculation.

\section{Excitation of multiphonon states}

\subsection{The coupling interaction}

The coupling interaction for the nuclear excitation $i\longrightarrow f$ in
a semi-classical calculation for an electric $\left( \pi =E\right) $, or
magnetic $\left( \pi =M\right) $, multipolarity, is given by (eqs. (6-7) of
ref. \cite{Ber98})

\begin{equation}
W_C=\frac{V_C}{\epsilon _0}=\sum_{\pi \lambda \mu }W_{\pi \lambda \mu
}\left( \tau \right) \;,  \label{wc1}
\end{equation}
where

\begin{equation}
W_{\pi \lambda \mu }\left( \tau \right) =\left( -1\right) ^{\lambda +1}\frac{
Z_1e}{\hbar vb^\lambda }\ \frac 1\lambda\ \sqrt{\frac{2\pi }{\left( 2\lambda
+1\right) !!}}Q_{\pi \lambda \mu }\left( \xi ,\tau \right) {\cal M}\left(
\pi \lambda ,\mu \right) \;.  \label{wc2}
\end{equation}

\noindent In this expression $b$ is the impact parameter, $\tau =\gamma vt/b$
is a dimensionless time variable with $\gamma =\left( 1-\beta
^2\right)^{-1/2} $ and $\beta =v/c$ being the usual relativistic parameters.
The energy scale is set by $\epsilon _0=\gamma \hbar v/b$ and the quantity $%
Q_{\pi \lambda \mu }\left( \xi ,\tau \right) $, with $\xi$ defined as the
adiabatic parameter $\xi =\xi _{if}=\left( E_f-E_i\right) /\epsilon _0$,
depends exclusively on the properties of the projectile-target relative
motion. The multipole operators acting on the intrinsic degrees of freedom
are, as usual,

\begin{equation}
{\cal M}(E\lambda ,\mu )=\int d^3r \ \rho ({\bf r})\ r^\lambda \ Y_{1\mu }( 
{\bf r})\ ,  \label{ME1}
\end{equation}
and

\begin{equation}
{\cal M}(M1,\mu )=-{\frac i{2c}}\ \int d^3r\ {\bf J}({\bf r}).{\bf L}\left(
rY_{1\mu }\right) \ ,  \label{MM1}
\end{equation}
We treat the excitation problem by the method of Alder and Winther\cite{AW}.
We solve a time-dependent Schr\"odinger equation for the intrinsic degrees
of freedom in which the time dependence arises from the projectile-target
motion, approximated by the classical trajectory. For relativistic energies,
a straight line trajectory is a good approximation. The wave function is
expanded in the relevant eigenstates of the nuclear Hamiltonian, $\{\mid
k\rangle ;\ k=1,N\}$, $N$ being the number of relevant intrinsic states
included in the coupled-channels (CC) problem.

In order to simplify the expression for the coupled equations we define the
dimensionless parameter $\Gamma _{kj}^{(\lambda )}$ by the relation

\begin{equation}
\Gamma _{kj}^{(\lambda )}=\left( -1\right) ^{\lambda +1}\frac{Z_1e}{\hbar 
{\rm v}b^\lambda }\frac 1\lambda \sqrt{\frac{2\pi }{\left( 2\lambda
+1\right) !!}}\;{\cal M}_{kj}(E\lambda).
\end{equation}

\noindent The coupled channels equations can then be written in the form 
\cite{Ber98}

\begin{equation}
\frac{da_k(\tau )}{d\tau }=-i\sum_{r=1}^N\ \sum_{\pi \lambda \mu }Q_{\pi
\lambda \mu }(\xi _{kj},\tau )\Gamma _{kj}^{\left( \lambda \right) }\;\exp
\left( i\xi _{kj}\tau \right) \;a_j(\tau )\;.  \label{ats1}
\end{equation}

\noindent In what follows we concentrate on the $E1$ excitation mode. In
this case, we have

\begin{equation}
Q_{E10}(\xi ,\tau )=\gamma \sqrt{2}\left[ \tau \phi ^3(t)-i\xi \left( \frac{%
{\rm v}}c\right) ^2\phi (t)\right] \;;\;\;\;\;\;Q_{E1\pm 1}(\xi ,\tau )=\mp
\phi ^3(\tau )\;,  \label{QE1}
\end{equation}
where $\phi (\tau )=\left( 1+\tau ^2\right) ^{1/2}.$

The excitation probability $P_j(b)$ of an intrinsic state $\mid j \rangle $
in a collision with impact parameter $b$ is obtained from the amplitudes $%
a_j(\tau )$ at asymptotically large times, in terms of an average over the
initial and a sum over the final magnetic quantum numbers. The cross section
is then obtained from the classical expression 
\begin{equation}
\sigma _j=2\pi \ \int P_j(b)\ T(b)\ b\;db\;.
\end{equation}
where the impact parameter dependent transmission factor $T(b)$ accounts for
absorption \cite{Ber98}.

\subsection{The 1-phonon states}

Consider the excitation of a nucleus with ground state spin zero (any
even-even nucleus) and isospin $T_{0}$, with its $3$-component $T_{3}=T_{0}$%
. In terms of the relevant quantum numbers, these states are written as: $%
|j>=|E_{j}^{(n)};J_{j}M_{j};T_{j}T_{3j}>$, where $n$ is the number of
phonons, $E$ the energy, $J_{j}$ and $M_{j}$ are respectively the spin and
its z-component quantum numbers, and $T_{j}$ and $T_{3j}$ are the isospin
and its third component. Note that due to charge conservation, all states
have $T_{3j}=T_{0}$. The ground state and the 1-phonon states are given in
the Table II below.

\bigskip

\begin{center}
\begin{tabular}{llll}
\hline
\ \ n\ \ \ \ \ \ \ \ \  & E\ \ \ \ \ \ \ \ \ \ \ \ \ \ \ \ \ \ \ \ \ \ \ \ \
\  & $J^{\pi }\;\;\;\;\;\;\;\;\;\;\;$ & T\ \ \ \ \ \ \ \ \ \ \ \ \ \ \ \  \\ 
\hline
$\;\;\;0$ & $0$ & 0$^{+}$ & T$_{0}$ \\ \hline
$\;\;\;1$ & $E_{GDR}$ & 1$^{-}$ & T$_{0}$ \\ 
& $E_{GDR}+\Delta _{T_{0}+1}^{(1)}$ & 1$^{-}$ & T$_{0}+1$ \\ \hline
\end{tabular}
\end{center}

\medskip

\begin{center}
{\bf Table II.} {\it Ground-state and one-phonon states with angular
momentum and isospin dependence. }
\end{center}

\noindent The energy splitting $\Delta _{T_0}^{(1)}$, which is assumed to
depend exclusively on isospin, is given by eq. \ref{fall1}.

In order to calculate the matrix elements ${\cal M}_{ki}\left( \pi \lambda
,\mu \right) $ between initial, $i$, and final states, $k$, we use the
Wigner-Eckart theorem in spin-isospin space and (except for the energy
dependence) assume that the {\it reduced matrix elements are isospin
independent}. We get

\begin{eqnarray}
{\cal M}_{ki}^{(10)}\left( \pi \lambda ,\mu \right) &=&\left\langle
E_{T_k}^{(1)};1M_k;T_kT_0\right| {\cal M}\left( E1,\mu \right) \left|
E^{\left( 0\right) };00;T_0T_0\right\rangle  \nonumber \\
&=&\left( -1\right) ^{1-M_f+T_k-T_0}\left( 
\begin{tabular}{lll}
$1$ & $1$ & $0$ \\ 
$-M_k$ & $\mu $ & $0$%
\end{tabular}
\right) \left( 
\begin{tabular}{lll}
$T_k$ & $1$ & $T_0$ \\ 
$-T_0$ & $0$ & $T_0$%
\end{tabular}
\right)  \nonumber \\
&&\times \langle 1||{\cal M}\left( E1\right) ||0\rangle \;.  \label{mfi1}
\end{eqnarray}

The value of the reduced matrix element $\langle 1||{\cal M}\left( E1\right)
||0\rangle $ can be extracted from the energy-weighted dipole sum-rule

\begin{equation}
S =\sum_{M_kT_k}\left( E_k^{(1)}-E^{(0)}\right) \left| \left\langle
E_{T_k}^{(1)};1M_k;T_kT_0\right| {\cal M}\left( E1,\mu \right) \left|
E^{\left( 0\right) };00;T_0T_0\right\rangle \right| \; =\frac 3{4\pi }\frac{
\hbar ^2}{2m_N}\frac{NZ}Ae^2\;.  \label{s1}
\end{equation}

\noindent Inserting the matrix-elements of eq. \ref{mfi1} in eq. \ref{s1},
we get

\begin{equation}
S = \sum_{T_k}\left( E_k^{(1)}-E^{(0)}\right) \langle 1||{\cal M}\left(
E1\right) ||0\rangle ^2\left( 
\begin{tabular}{lll}
$T_k$ & $1$ & $T_0$ \\ 
$-T_0$ & $0$ & $T_0$%
\end{tabular}
\right) ^2 \times \sum_{M_k}\left( 
\begin{tabular}{lll}
$1$ & $1$ & $0$ \\ 
$-M_k$ & $\mu $ & $0$%
\end{tabular}
\right) ^2\;.  \label{s2}
\end{equation}

\noindent Using the relation \cite{Ed60},

\begin{equation}
\sum_{M_k}\left( 
\begin{tabular}{lll}
$1$ & $1$ & $0$ \\ 
-$M_k$ & $\mu $ & $0$%
\end{tabular}
\right) ^2=\left( 
\begin{tabular}{lll}
$1$ & $1$ & $0$ \\ 
$-\mu $ & $\mu $ & $0$%
\end{tabular}
\right) ^2=\frac 13\;,
\end{equation}

\noindent the sum rule takes the form 
\begin{equation}
S =\frac{\langle 1||{\cal M}\left( E1\right) ||0\rangle ^2}3\Biggl\{\left( 
\begin{tabular}{lll}
$T_0$ & $1$ & $T_0$ \\ 
$-T_0$ & $0$ & $T_0$%
\end{tabular}
\right) ^2E_{GDR} \; +\left( 
\begin{tabular}{lll}
$T_0+1$ & $1$ & $T_0$ \\ 
$-T_0$ & $0$ & $T_0$%
\end{tabular}
\right) ^2\left( E_{GDR}+\Delta _{T_0+1}^{(1)}\right) \Biggr\}\;.
\label{sum1}
\end{equation}

\noindent Using the explicit forms of the Wigner coefficients \cite{Ed60},

\begin{eqnarray}
\left( 
\begin{tabular}{lll}
$T_0$ & $1$ & $T_0$ \\ 
$-T_0$ & $0$ & $T_0$%
\end{tabular}
\right) &=&\left[ \frac{T_0}{\left( T_0+1\right) \left( 2T_0+1\right) }
\right] ^{1/2}\;;\;\;  \nonumber \\
\;\;\;\;\;\left( 
\begin{tabular}{lll}
$T_0+1$ & $1$ & $T_0$ \\ 
$-T_0$ & $0$ & $T_0$%
\end{tabular}
\right) &=&-\left[ \frac 1{\left( T_0+1\right) \left( 2T_0+3\right) }\right]
^{1/2}\;,  \label{sum2}
\end{eqnarray}
\noindent we obtain the reduced matrix element

\begin{equation}
\langle 1||{\cal M}\left( E1\right) ||0\rangle =\sqrt{\frac{3S}{%
E_{GDR}F(T_0) }}\;,  \label{sum3}
\end{equation}
with

\begin{equation}
F\left( T_0\right) =\frac 1{T_0+1}\left\{ \frac{T_0}{2T_0+1}+\left[ 1+\frac{%
\Delta _{T_0+1}^{(1)}}{E_{GDR}}\right] \frac 1{2T_0+3}\right\} \;.
\label{ft}
\end{equation}

\subsection{The two-phonon states}

The two-phonon states must be symmetric with respect to the exchange of the
two phonons in spin and isopin. This symmetry can be tracked by using the
coupling scheme $\left| \left(1_{1}1_{2}\right) \Im T_{0}; T_{f}
T_{0}\right\rangle$. One has to distinguish the two cases $\Im =0$ (isospin
even) and $\Im =1$ (isospin odd). The states which are spin-isospin
symmetric correspond then to the combinations

\begin{equation}
\left(I_f,\Im\right)=\left(0,0\right);\; \left(0,2\right);\;
\left(2,0\right);\; \left(2,2\right);\; \left(1,1\right).
\end{equation}

\noindent The two-phonon states are represented as $|j>=|E_{j}^{(2)};
J_{j}M_{j};\left( 1_{1}1_{2}\right) \Im T_{0}\ T_{f}T_{0}>$. We list the
main quantum numbers of the two-phonon states in table III.

\bigskip

\begin{center}
\begin{tabular}{llll}
\hline
$\;\;\;\Im \;\;\;\;\;\;$ & E\ \ \ \ \ \ \ \ \ \ \ \ \ \ \ \ \ \ \ \ \ \ \ \
\ \  & J$^{\pi }\;\;\;\;\;\;\;$ & T\ \ \ \ \ \ \ \ \ \ \ \  \\ \hline
$\;\;\;0$ & $E_{DGDR}$ & 0$^{+}$ & T$_{0}$ \\ 
& $E_{DGDR}$ & 2$^{+}$ & T$_{0}$ \\ \hline
$\;\;\;1$ & $E_{DGDR}$ & 1$^{+}$ & T$_{0}$ \\ 
& $E_{DGDR}+\Delta _{T_{0}+1}^{(2)}$ & 1$^{+}$ & T$_{0}+1$ \\ \hline
$\;\;\;2$ & $E_{DGDR}$ & 0$^{+}$ & T$_{0}$ \\ 
& $E_{DGDR}$ & 2$^{+}$ & T$_{0}$ \\ 
& $E_{DGDR}+\Delta _{T_{0}+1}^{(2)}$ & 0$^{+}$ & T$_{0}+1$ \\ 
& $E_{DGDR}+\Delta _{T_{0}+1}^{(2)}$ & 2$^{+}$ & T$_{0}+1$ \\ 
& $E_{DGDR}+\Delta _{T_{0}+2}^{(2)}$ & 0$^{+}$ & T$_{0}+2$ \\ 
& $E_{DGDR}+\Delta _{T_{0}+2}^{(2)}$ & 2$^{+}$ & T$_{0}+2$ \\ \hline
\end{tabular}

{\bf Table III.} {\it Two-phonon states with angular momentum and isospin
dependence. The isospin dependence arises from the coupling of the phonon
isospins} $\Im =1_1\otimes 1_2$.
\end{center}

\bigskip

\noindent The isospin shifts $\Delta^{(2)}$ in the above table are given by
eqs. \ref{fall2} and \ref{fall3}.

To obtain the excitation amplitudes for the one phonon to two phonon
transitions we need to calculate the following matrix elements:

\begin{eqnarray}
{\cal M}_{fk}^{(21)}\left( E1,\mu \right) &=&\left\langle
E_{T_{f}}^{(2)};J_{f}M_{f};\left( 1_{1},1_{2}\;\right) \Im
T_{0};T_{f}T_{0}\right| {\cal M}\left( E1,\mu \right) \left| E^{\left(
1\right) };J_{k}M_{k};T_{k}T_{0}\right\rangle \;.  \nonumber \\
&&  \label{mfi2}
\end{eqnarray}

\noindent This can be done by changing the final state coupling scheme. We
use the relation $\left( \text{{\rm ref. \cite{Ed60},eq. 6.1.3, p.91}}%
\right) $

\begin{eqnarray}
\left| JM;\left( 1_11_2\right) \Im T_0;T_fT_0\right\rangle
&=&\sum_{T_l}\left( -1\right) ^{T_l+T_f}\sqrt{\left( 2\Im +1\right) \left(
2T_l+1\right) }  \nonumber \\
&&\times \left\{ 
\begin{array}{ccc}
1 & 1 & \Im \\ 
T_0 & T_f & T_l
\end{array}
\right\} \left| JM;\left( 1_2\left( 1_1T_0\right) T_l\right)
;T_fT_0\right\rangle \;.  \nonumber \\
&&  \label{sum4}
\end{eqnarray}

\noindent and obtain

\begin{eqnarray}
&&{\cal M}_{fk}^{(21)}\left( E1,\mu \right) =\sum_{T_{l}}\left( -1\right)
^{T_{l}+T_{f}}\sqrt{\left( 2\Im +1\right) \left( 2T_{l}+1\right) }\left\{ 
\begin{array}{ccc}
1 & 1 & \Im \\ 
T_{0} & T_{f} & T_{l}
\end{array}
\right\}  \nonumber \\
&&\times \left\langle E_{T_{f}}^{(2)};J_{f}M_{f};\left( 1_{2}\left(
1_{1}T_{0}\right) T_{l}\right) ;T_{f}T_{0}\right| {\cal M}\left( E1,\mu
\right) \left| E^{\left( 1\right) };J_{k}M_{k};T_{k}T_{0}\right\rangle \;. 
\nonumber \\
&&  \label{mfi3}
\end{eqnarray}

\noindent We next use the Wigner-Eckart theorem in spin-isospin space,
assuming that the reduced matrix element are spin and isospin independent
and vanish unless $T_k=T_l$. We get

\begin{eqnarray}
&&\left\langle E_{T_{f}}^{(2)};J_{f}M_{f};\left( 1_{2}\left(
1_{1}T_{0}\right) T_{l}\right) ;T_{f}T_{0}\right| {\cal M}\left( E1,\mu
\right) \left| E^{\left( 1\right) };J_{k}M_{k};\left( 1_{1}T_{0}\right)
T_{k}T_{0}\right\rangle =  \nonumber \\
&&\delta _{\left( T_{l},T_{k}\right) }\cdot \left( -1\right)
^{J_{f}-M_{f}+T_{f}-T_{0}}\left( 
\begin{tabular}{lll}
$1$ & $1$ & $0$ \\ 
$-M_{f}$ & $\mu $ & $0$%
\end{tabular}
\right) \left( 
\begin{tabular}{lll}
$T_{f}$ & $1$ & $T_{k}$ \\ 
$-T_{0}$ & $0$ & $T_{0}$%
\end{tabular}
\right) \langle 2||{\cal M}\left( E1\right) ||1\rangle \;.  \nonumber \\
&&  \label{sum5}
\end{eqnarray}

\noindent Using $\langle 2||{\cal M}\left( E1\right) ||1\rangle =\sqrt{2}
\langle 1||{\cal M}\left( E1\right) ||0\rangle $ in eq. \ref{sum5}, eq. \ref
{mfi3} becomes

\begin{eqnarray}
{\cal M}_{fk}^{(21)}\left( E1,\mu \right) &=&\sqrt{2}\left( -1\right)
^{J_{f}-M_{f}+2T_{f}-T_{0}+T_{k}}\langle 1||{\cal M}\left( E1\right)
||0\rangle \sqrt{\left( 2\Im +1\right) \left( 2T_{k}+1\right) }  \nonumber \\
&&\times \left( 
\begin{tabular}{lll}
$1$ & $1$ & $0$ \\ 
$-M_{f}$ & $\mu $ & $0$%
\end{tabular}
\right) \left( 
\begin{tabular}{lll}
$T_{f}$ & $1$ & $T_{k}$ \\ 
$-T_{0}$ & $0$ & $T_{0}$%
\end{tabular}
\right) \left\{ 
\begin{array}{ccc}
1 & 1 & \Im \\ 
T_{0} & T_{f} & T_{k}
\end{array}
\right\} \;.  \label{mfi4}
\end{eqnarray}

\section{Applications}

In this section, we apply our results to the excitation of one- and
two-phonons states in $^{208}$Pb and $^{48}$Ca target nuclei, in collisions
with relativistic $^{208}$Pb projectiles. Before we present the numerical
results, we rewrite the matrix elements so that they can be used as input to
the coupled-channel code RELEX for Coulomb excitation \cite{Ber98}. They
become ``effective reduced matrix elements'' that incorporate the isospin 
dependence 
of ${\cal M}_{ki}^{(10)}\left( \pi \lambda ,\mu \right) $ (eq.~\ref{mfi1}). 
Namely, 
\begin{equation}
\langle 1||{\cal M}\left( E1\right) ||0\rangle _{ki}^{(eff)}=\left(
-1\right) ^{T_{k}-T_{0}}\left( 
\begin{tabular}{lll}
$T_{k}$ & $1$ & $T_{0}$ \\ 
$-T_{0}$ & $0$ & $T_{0}$%
\end{tabular}
\right) \langle 1||{\cal M}\left( E1\right) ||0\rangle \;,  \label{mred1}
\end{equation}
while for equation \ref{mfi4} it is more convenient to define

\begin{equation}
\langle 2||{\cal M}\left( E1\right) ||1\rangle _{fk}^{(eff)}=\sqrt{2}\left(
-1\right) ^{2T_{f}-T_{0}+T_{k}}\langle 1||{\cal M}\left( E1\right)
||0\rangle \sqrt{\left( 2\Im +1\right) \left( 2T_{k}+1\right) }\left( 
\begin{tabular}{lll}
$T_{f}$ & $1$ & $T_{k}$ \\ 
$-T_{0}$ & $0$ & $T_{0}$%
\end{tabular}
\right) \left\{ 
\begin{array}{ccc}
1 & 1 & \Im  \\ 
T_{0} & T_{f} & T_{k}
\end{array}
\right\} \;.  \label{mred2}
\end{equation}

\noindent The modified reduced matrix-elements for one- and two-phonon
excitations are presented in tables IV and V, in the cases of $^{208}$Pb and 
$^{48}$Ca, respectively. For comparison the original reduced matrix-elements
are also given in the last column.

\bigskip

\begin{center}
\begin{tabular}{llllllll}
\hline
$E_{k}\;\;\;\;\;\;$ & $E_{f}\;\;\;\;\;\;$ & $J_{k}\;\;\;\;\;\;$ & $%
J_{f}\;\;\;\;\;\;$ & $T_{k}\;\;\;\;\;\;$ & $T_{f}\;\;\;\;\;\;$ & $\Im
\;\;\;\;\;\;$ & $\langle n+1||{\cal M}\left( E1\right) ||n\rangle
_{fk}^{(eff)}\;\;\;\;\;\;$ \\ \hline
0 & 13.5 & 0 & 1 & 22 & $22$ &  & 7.16\ \ \ \ (7.39) \\ 
0 & 20.1 & 0 & 1 & 22 & $23$ &  & 1.53 \\ 
13.5 & 27 & 1 & 0 & 22 & 22 & 0 & $-5.84$\ \ (10.45) \\ 
13.5 & 27 & 1 & 2 & 22 & 22 & 0 & $-5.84$\ \ (10.45) \\ 
13.5 & 27 & 1 & 1 & 22 & 22 & 1 & 0.318 \\ 
13.5 & 33.6 & 1 & 1 & 22 & 23 & 1 & $-1.46$ \\ 
13.5 & 27 & 1 & 0 & 22 & 22 & 2 & 8.25 \\ 
13.5 & 27 & 1 & 2 & 22 & 22 & 2 & 8.25 \\ 
13.5 & 33.6 & 1 & 0 & 22 & 23 & 2 & 1.53 \\ 
13.5 & 33.6 & 1 & 2 & 22 & 23 & 2 & 1.53 \\ 
13.5 & 40.8 & 1 & 0 & 22 & 24 & 2 & 0 \\ 
13.5 & 40.8 & 1 & 2 & 22 & 24 & 2 & 0 \\ 
20.1 & 27 & 1 & 0 & 23 & 22 & 0 & 1.25 \\ 
20.1 & 27 & 1 & 2 & 23 & 22 & 0 & 1.25 \\ 
20.1 & 27 & 1 & 1 & 23 & 22 & 1 & 1.49 \\ 
20.1 & 33.6 & 1 & 1 & 23 & 23 & 1 & $-6.85$ \\ 
20.1 & 27 & 1 & 0 & 23 & 22 & 2 & 0.824 \\ 
20.1 & 27 & 1 & 2 & 23 & 22 & 2 & 0.824 \\ 
20.1 & 33.6 & 1 & 0 & 23 & 23 & 2 & $-6.56$ \\ 
20.1 & 33.6 & 1 & 2 & 23 & 23 & 2 & $-6.56$ \\ 
20.1 & 40.8 & 1 & 0 & 23 & 24 & 2 & $-2.83$ \\ 
20.1 & 40.8 & 1 & 2 & 23 & 24 & 2 & $-2.83$ \\ \hline
\end{tabular}
\end{center}

\medskip

\begin{center}
{\bf Table IV.} {\it Reduced matrix elements for transitions from the GS to
one-phonon states and also from one to two-phonon states, for $^{208}Pb$ in
units of fm.$e.$ The numbers within parentheses in the last column are the
corresponding reduced matrix elements without consideration of isospin. To
avoid ambiguities in the two-phonon final states, we also indicate the
isospin of the coupled-phonon pair.}
\end{center}

\bigskip

\begin{center}
\begin{tabular}{llllllll}
\hline
$E_{k}\;\;\;\;\;\;$ & $E_{f}\;\;\;\;\;\;$ & $J_{k}\;\;\;\;\;\;$ & $%
J_{f}\;\;\;\;\;\;$ & $T_{k}\;\;\;\;\;\;$ & $T_{f}\;\;\;\;\;\;$ & $\Im
\;\;\;\;\;\;$ & $\langle n+1||{\cal M}\left( E1\right) ||n\rangle
_{fk}^{(eff)}\;\;\;\;\;\;$ \\ \hline
0 & 19.2 & 0 & 1 & 4 & $4$ &  & 2.66\ \ \ \ (3.00) \\ 
0 & 25.4 & 0 & 1 & 4 & $5$ &  & 1.2 \\ 
19.2 & 38.4 & 1 & 0 & 4 & 4 & 0 & $-2.17$ \ (4.24) \\ 
19.2 & 38.4 & 1 & 2 & 4 & 4 & 0 & $-2.17$ \ (4.24) \\ 
19.2 & 38.4 & 1 & 1 & 4 & 4 & 1 & 0.594 \\ 
19.2 & 44.6 & 1 & 1 & 4 & 5 & 1 & $-1.08$ \\ 
19.2 & 38.4 & 1 & 0 & 4 & 4 & 2 & 3.01 \\ 
19.2 & 38.4 & 1 & 2 & 4 & 4 & 2 & 3.01 \\ 
19.2 & 44.6 & 1 & 0 & 4 & 5 & 2 & 1.31 \\ 
19.2 & 44.6 & 1 & 2 & 4 & 5 & 2 & 1.31 \\ 
19.2 & 53.4 & 1 & 0 & 4 & 6 & 2 & 0 \\ 
19.2 & 53.4 & 1 & 2 & 4 & 6 & 2 & 0 \\ 
25.4 & 38.4 & 1 & 0 & 5 & 4 & 0 & 1.09 \\ 
25.4 & 38.4 & 1 & 2 & 5 & 4 & 0 & 1.09 \\ 
25.4 & 38.4 & 1 & 1 & 5 & 4 & 1 & 1.19 \\ 
25.4 & 44.6 & 1 & 1 & 5 & 5 & 1 & $-2.15$ \\ 
25.4 & 38.4 & 1 & 0 & 5 & 4 & 2 & 0.548 \\ 
25.4 & 38.4 & 1 & 2 & 5 & 4 & 2 & 0.548 \\ 
25.4 & 44.6 & 1 & 0 & 5 & 5 & 2 & $-1.76$ \\ 
25.4 & 44.6 & 1 & 2 & 5 & 5 & 2 & $-1.76$ \\ 
25.4 & 53.4 & 1 & 0 & 5 & 6 & 2 & $-1.92$ \\ 
25.4 & 53.4 & 1 & 2 & 5 & 6 & 2 & $-1.92$ \\ \hline
\end{tabular}
\end{center}

\medskip

\begin{center}
{\bf Table V.} {\it Same as Table V, for $^{48}$Ca}
\end{center}

The calculated excitation cross sections for one and two phonon states in
the $^{208}$Pb (650 MeV $\cdot $ A) + $^{208}$Pb collision are given in
table VI. They are also plotted in figure 1$a$, as a function of the
excitation energy. In this figure, the dominant spin and parity is indicated
in each case. For a comparison, corresponding results neglecting isospin are
given within parentheses in table VI and are shown in figure 1$b$. If
isospin is neglected, only $T=22$ (the ground state isospin of $^{208}$Pb)
states are populated. Namely, the GDR state at 13.5 MeV and $0^{+}$ and $%
2^{+}$ DGDR states at 27.0 MeV. With the inclusion of isospin, about 3 \% 
of the GDR cross section is associated with the population of the $T=23$ 
state at 20.1 MeV, as can be seen in table VI and in figure 1$a$.  This
corresponds to the exhaustion of about 6.3 \% of the GDR sum rule.
The consequences of the isospin degree
of freedom on the DGDR population are more important. Although most of the
cross section remains associated with the $T=22$, $J^{\pi }=0^{+}$ and $2^{+}
$ states at 27.0 MeV, 7 \% of the total DGDR cross section then arises from
the population of the $T=23$ states at 33.6 MeV, of which over 90 \%
corresponds to the non-natural parity $J^{\pi }=1^{+}$ state. Thus, 6 \% of
the DGDR cross section goes to the excitation of the $J^{\pi }=1^{+}$ state,
which would be forbidden in the usual harmonic oscillator picture (without
isospin).  On the other hand, some calculations using RPA descriptions
of the giant resonances find non-vanishing population of such states.  
Lanza {\sl et al.} \cite{Vo95} find negligible populations while Bertulani 
{\sl et al.} \cite{BPV96} get about half of that of the present work. 
This result could be traced back to the fact that their second
order transitions cancel exactly, so that $1^{+}$ states can only be
reached through higher order coupled channel processes.

Table VII and figure 2$a$ give similar results for the excitation of GDR and
DGDR states in $^{48}$Ca, in the collision $^{208}$Pb (650 MeV $\cdot $ A) + 
$^{48}$Ca. Corresponding results neglecting isospin are given in the same
way as above.  For this system, isospin plays a more important role due to
the lower isospin quantum number ($T=4$) of the $^{48}$Ca ground state. 
In this case, the dominant $T=4$ GDR state at 19.2 MeV loses more than 
10 \% of its cross section to the $T=5$ $1^{-}$ state at 25.4 MeV, which
exhausts  about 21 \% of the GDR sum rule.
Moreover, the DGDR cross section is very much affected by isospin.
Figure 2$a$ indicates that the cross section for $T=5$ DGDR states at 44.6
MeV reaches 32~\% of that for the dominant $T=4$ DGDR states at 33.6 MeV. It
is also important to discuss the $J^{\pi }$ distribution of the DGDR cross
section. Since over 95 \% of the $T=5$ DGDR cross section corresponds to $%
J^{\pi }=1^{+}$, the population of states with this spin and parity is
rather large. 

\begin{center}
\begin{tabular}{llllll}
\hline
\ \ \ \  & E (MeV)\ \  & J$^{\pi }\;\;\;\;\;\;$ & T\ \ \ \ \ \ \ \  & $\Im
\;\;\;\;\;\;$ & $\sigma $ (mb) \\ \hline
GDR & 13.5 & 1$^{-}$ & 22 &  & 2240\ \ \ \ \ \ (2597) \\ 
& 20.1 & 1$^{-}$ & 23 &  & 59.86 \\ \hline
DGDR & 27.0 & 0$^{+}$ & 22 & 0 & 4.489\ \ \ \ \ (15.80)\  \\ 
& 27.0 & 2$^{+}$ & 22 & 0 & 31.16 \ \ \ \ (117.5) \\ 
& 27.0 & 1$^{+}$ & 22 & 1 & 0.8608 \\ 
& 33.6 & 1$^{+}$ & 23 & 1 & 8.373 \\ 
& 27.0 & 0$^{+}$ & 22 & 2 & 10.85 \\ 
& 27.0 & 2$^{+}$ & 22 & 2 & 69.89 \\ 
& 33.6 & 0$^{+}$ & 23 & 2 & 0.1257 \\ 
& 33.6 & 2$^{+}$ & 23 & 2 & 0.3732 \\ 
& 40.8 & 0$^{+}$ & 24 & 2 & 0.0237 \\ 
& 40.8 & 2$^{+}$ & 24 & 2 & 0.0874 \\ \hline
\end{tabular}

{\bf Table VI.} {\it Excitation cross sections of one and two phonon states
in the collision }$^{208}Pb$ {\it (640\ }$\cdot \;${\it A MeV) +\ }$^{208}Pb$%
.

\bigskip

\begin{tabular}{llllll}
\hline
\ \ \ \  & E (MeV)\ \  & J$^{\pi }\;\;\;\;\;\;$ & T\ \ \ \ \ \ \ \  & $\Im
\;\;\;\;\;\;$ & $\sigma $ (mb) \\ \hline
GDR & 19.2 & 1$^{-}$ & 4 &  & 318.5\ \ \ \ \ \ (405.5) \\ 
& 25.4 & 1$^{-}$ & 5 &  & 33.11 \\ \hline
DGDR & 38.4 & 0$^{+}$ & 4 & 0 & 0.2842\ \ \ \ \ (2.137)\  \\ 
& 38.4 & 2$^{+}$ & 4 & 0 & 0.4453 \ \ \ \ (3.613) \\ 
& 38.4 & 1$^{+}$ & 4 & 1 & 0.6695 \\ 
& 44.6 & 1$^{+}$ & 5 & 1 & 1.275 \\ 
& 38.4 & 0$^{+}$ & 4 & 2 & 1.117 \\ 
& 38.4 & 2$^{+}$ & 4 & 2 & 1.640 \\ 
& 44.6 & 0$^{+}$ & 5 & 2 & 0.0155 \\ 
& 44.6 & 2$^{+}$ & 5 & 2 & 0.0366 \\ 
& 53.4 & 0$^{+}$ & 6 & 2 & 0.0268 \\ 
& 53.4 & 2$^{+}$ & 6 & 2 & 0.0375 \\ \hline
\end{tabular}

{\bf Table VII.} {\it Same as Table VI, for the excitation of $^{48}$Ca in
the collision }$^{208}Pb$ {\it (640\ }$\cdot \;${\it A MeV) +\ }$^{48}Ca$.
\end{center}

\bigskip

\section{Conclusions}

In this paper isospin is taken into account in the excitation of the double
giant dipole resonance. We have used a semiclassical coupled-channels 
formalism to calculate excitation probabilities and cross sections for 
the collisions $^{208}{\rm Pb}\ (640\ \cdot\  A\ {\rm MeV})\ +\ 
^{208}{\rm Pb}$ and $^{208}{\rm Pb}\ (640\ \cdot \ A\ {\rm MeV})\ +\ 
^{48}{\rm Ca}$. We have assumed that the eletromagnetic matrix elements are
isospin independent and adopted the isospin splitting of energy levels
as given by Aky\"uz and Fallieros \cite{Fa71}. It has been shown that 
isospin leads to a redistribution of the strengths of the electromagnetic 
matrix elements such that the probability for Coulomb excitation of 
$J^{\pi }=1^{+}$ DGDR states is enhanced.  This enhancement depends on 
3J and 6J coefficients in isospin space so that it becomes more relevant 
for nuclei with low ground state isospin.  Consequently,  this effect 
turned out to be much stronger in the excitation of $^{48}{\rm 
Ca}$ than in the excitation of $^{208}{\rm Pb}$.  One should then expect 
that isospin splitting should contribute
to make de DGDR broader, particularly for nuclei with low ground state
isospin.  This result suggests that our formalism should be especially 
relevant to study the excitation cross sections of a family of isotopes 
with charge numbers varying from the proton to the neutron dripline. Study 
along these directions is underway.

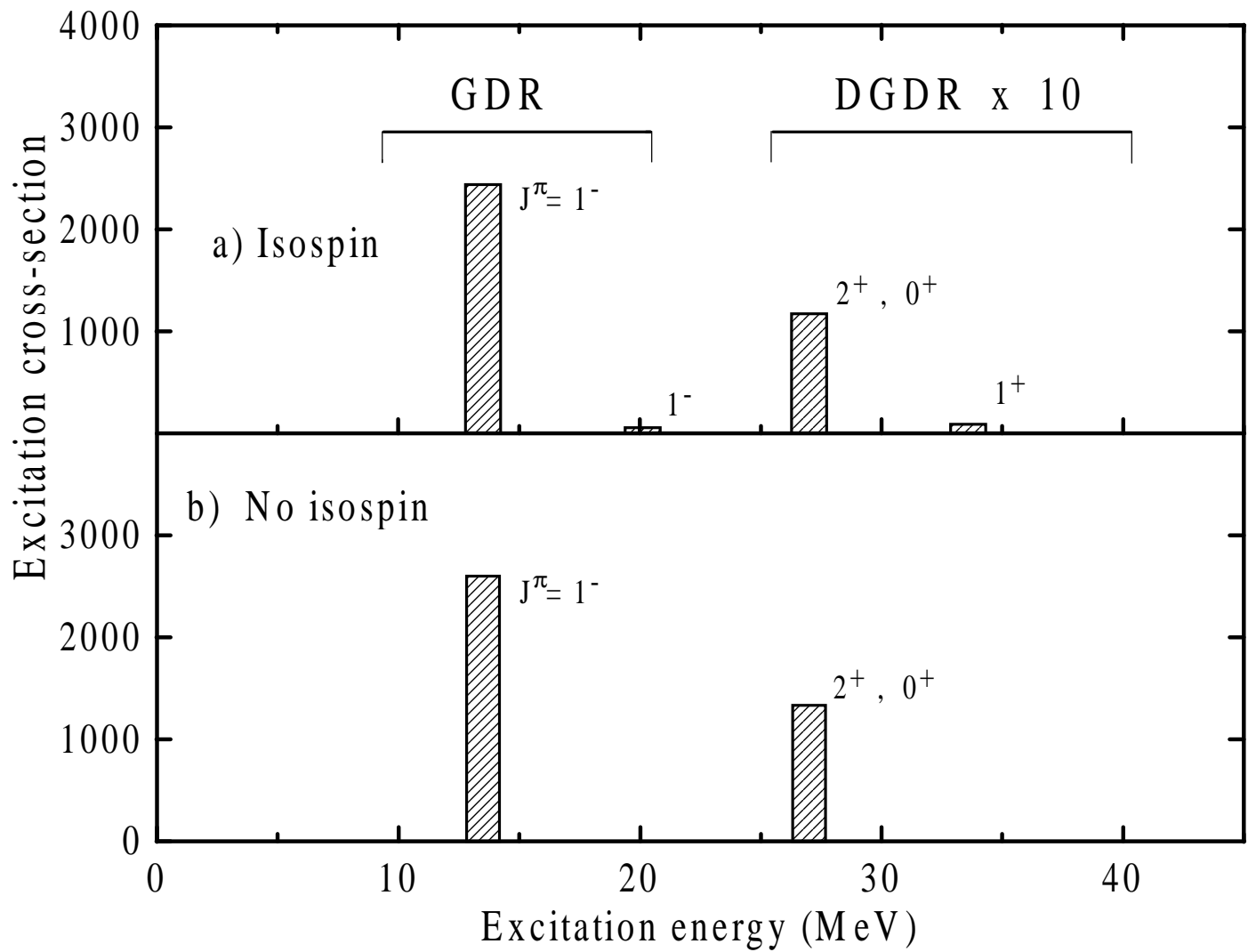
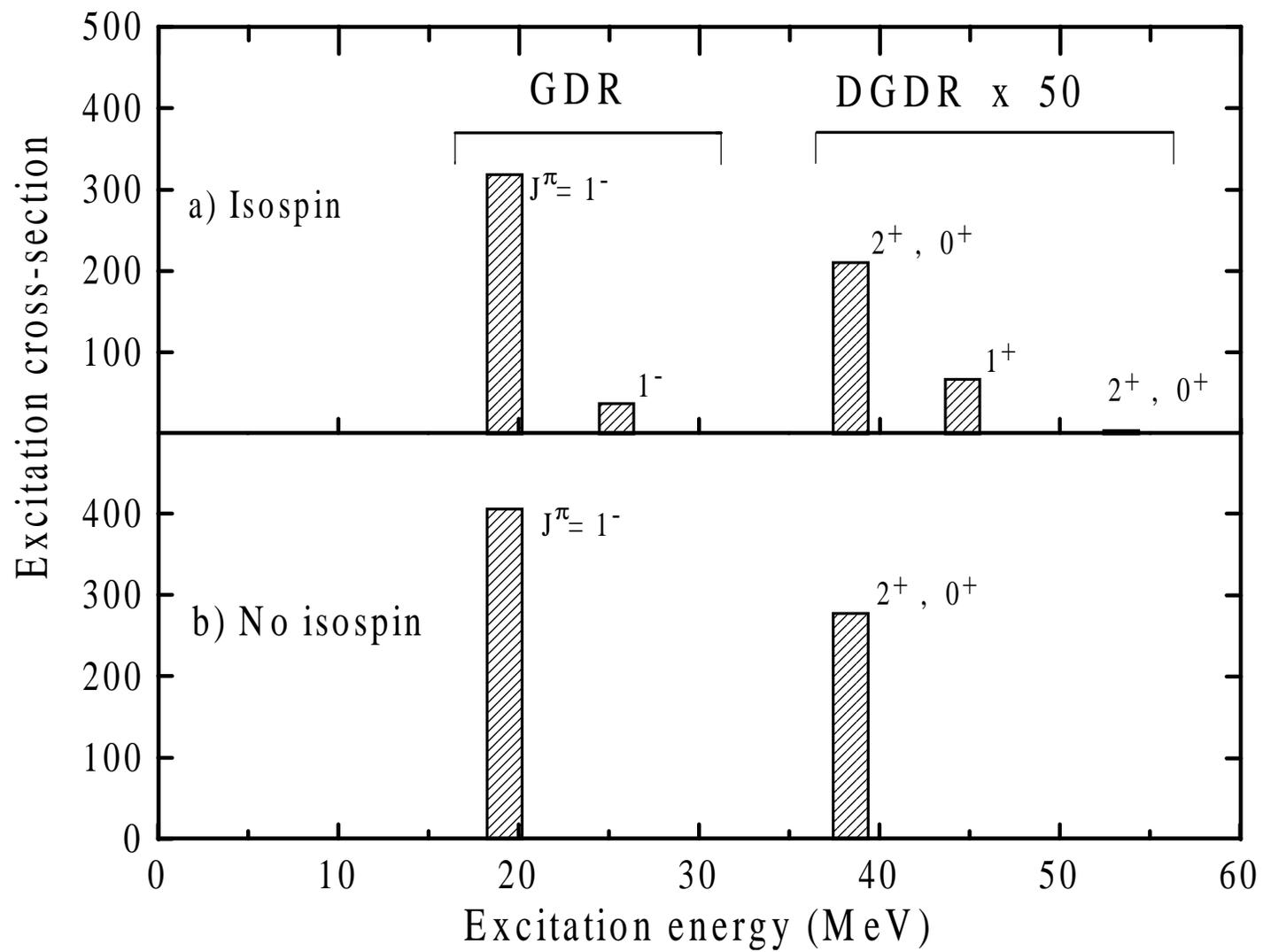
\end{document}